\documentclass[epj,ctagsplt]{svjour}
\usepackage{epsfig}
\usepackage{amsmath}
\usepackage{amssymb}

\usepackage{color}

\begin{document}

\title{Random walks with imperfect trapping
        in the decoupled-ring approximation}
\author {Timo Aspelmeier\inst{1,2}
        \and J\'er\^ome Magnin\inst{1,3}
        \and Willi Graupner\inst{1,4}
        \and Uwe C. T\"auber\inst{1}            }
\institute{Department of Physics, Virginia Polytechnic Institute and State
    University, Blacksburg, VA 24061-0435, USA
        \and Present address: Department of Physics and Astronomy,
           University of Manchester, Oxford Road, Manchester M13 9PL, UK
    \and Present address: GeneProt Inc., 2 rue Pr\'e-de-la-Fontaine,
    CH-1217 Meyrin, Switzerland
    \and Present address: austriamicrosystems AG, A-8141 Schloss Premst\"atten,
            Austria }

\date{\today}

\abstract{We investigate random walks on a lattice with imperfect traps.
In one dimension, we perturbatively compute the survival probability by
reducing the problem to a particle diffusing on a closed ring containing
just one single trap.
Numerical simulations reveal this solution, which is exact in the limit
of perfect traps, to be remarkably robust with respect to a significant
lowering of the trapping probability.
We demonstrate that for randomly distributed traps, the long-time
asymptotics of our result recovers the known stretched exponential decay.
We also study an anisotropic three-dimensional version of our model,
where for sufficiently large transverse diffusion the system is described
by the mean-field kinetics.
We discuss possible applications of some of our findings to the decay of
excitons in semiconducting organic polymer materials, and emphasize the
crucial influence of the spatial trap distribution on the kinetics.}


\maketitle


\section{Introduction}
\label{Intro}

Random walks in disordered media~\cite{Kehr-review,Bouchaud-review}
form a class of statistical processes that have been widely employed
for the description of an impressive number of physical, chemical and
biological phenomena. A particularly interesting subclass is
constituted by Brownian motion in the presence of {\em quenched
disorder} in the medium. Depending on the nature of this disorder,
significant deviations from the results known for pure random walks
are observed. This is true for microscopic quantities (e.g. first
passage times, average distance from origin at time $t$, etc.) as well
as for macroscopic properties, such as the asymptotic time evolution
of the overall population of random walkers undergoing site-correlated
annihilation~\cite{Kehr-review,Hugues}.

In the present paper, we address the intermediate and long-time kinetics of
random walkers decaying through two simultaneous channels: (i) spontaneous
(`radioactive') decay, with no spatial or temporal correlations involved, and
(ii) a decay process induced by the presence of non-moving (quenched)
imperfect traps in the medium.

In the existing literature, the term `trap' has been used to describe
different types of disorder: In certain instances, the concept of a trap
has embodied potential wells in the medium, acting on a walker through
delaying its motion (the `valley model')~\cite{Kehr-review}.
In the present context, we refer to traps simply as sites where
particles may undergo spontaneous annihilation with a certain (fixed)
probability $0 < q \leq 1$.

Random walks with annihilation by trapping have already been the
object of considerable attention during the past two decades%
~\cite{Balagurov-Vaks,Dombetal,Donsker-Varadhan,%
Nieuwenhuizen1,Nieuwenhuizen2,Anlauf,WeiHav,%
Barkema,Gallos}, to name but a few which are of relevance for this work.
These works represent a fundamental and necessary step towards an
understanding of transport processes in partially absorbing media.
In biology, this applies, e.g., to the scattering of laser light in
tissues. In chemistry, microscopic theories of chemical reaction
kinetics necessitate to access quantities such as the probability
distribution of the nearest-neighbor distance of a diffusing molecule
to a (static or moving) trap, representing a reaction center.

In solid-state physics, the study of a whole set of different phenomena
in disordered systems relies heavily on the understanding of random
walks in the presence of partially absorbing traps:
Electron-hole recombination in amorphous solids, and chemical binding
by impurities of interstitial hydrogen atoms in metals represent two
common examples.
More indirectly, this problem is also related to self-attracting
polymers, and to the density of states of binary disordered systems,
see Ref.~\cite{Nieuwenhuizen1}.

We draw part of our motivation for this present study from a class of
organic semiconducting materials which has been the object of
considerable attention recently~\cite{ScienceBat}. Due to the nature
of their opto-electronic properties, thin films of organic molecules
working as semiconducting dio\-des~\cite{PhysicsWorld} are already
employed in high-performance electroluminescent displays~\cite{SPIE}.
For the initial experimental studies of excited states in these
materials, aiming at understanding their fundamental properties, light
is used for their generation --- in most cases the elementary
excitations take the form of a bound electron-hole
pair~\cite{SariSingapore}. These excitations behave as
pseudo-particles ({\em excitons}) which diffuse and interact in
varying and complex ways with the medium~\cite{Scheidler}, and
eventually annihilate by emitting photons or phonons. The latter,
non-radiative annihilation process is known to be mediated by chemical
defects~\cite{Yan}.

Experimental studies suggest that the physics of relaxation at work in
such excitable media depends in a rather tight and non-trivial way on a
variety of partially controllable parameters.
Among them, we quote the chemical nature of the material used, the way
it has been prepared and/or altered by techniques like photo-oxidation,
and the configurational structure at the microscopic level as well as
interactions between individual polymer chains.
These mutual interactions are markedly different in solid or liquid
solutions~\cite{rothberg95}, and amorphous or polycrystalline films,
respectively.
The detailed relaxation mechanisms, and their dependence on sample
properties are as yet incompletely understood, as far as their relation
to the elementary and fundamental processes involved are concerned.
We believe that the present, somewhat unsatisfactory situation calls for
an effort in investigating, e.g., through simple kinetic statistical
models, the role played by these different parameters.

In addition, the effectively one-dimensional (or at least strongly
directed, anisotropic) nature of exciton transport in organic
semiconductors renders simple mean-field theory approaches inadequate.
Rather, one expects marked fluctuation and correlation effects.
For the model studied here, these in turn should strongly depend on the
spatial defect distribution.
Thus, experiments probing the exciton kinetics in semiconducting
polymers promise to provide an excellent testing ground for a wide
variety of non-equilibrium statistical models.
Indeed, the dynamics of laser-induced excitons in N(CH$_3$)$_4$MnCl$_3$
(TMMC) polymer chains appears to be the best experimental realization to
date for diffusion-limited fusion processes $A + A \to A$, where (in an
intermediate time window) the predicted power law for the particle density
$n \sim t^{-1/2}$ has actually been observed unambiguously~\cite{Kroon93}.

The present paper is organized as follows: In Sec.~\ref{Model} we
provide a precise definition of our model of a random walk with
imperfect traps (inducing \textit{non-radiative} exciton
recombination) and spontaneous annihilation (or \textit{radiative}
recombination). Our main interest being the intermediate and long-time
decay of the overall population of particles, the following section
presents a calculation of the temporal behavior of this quantity in a
one-dimensional setup. It is known that the low dimensionality has a
strong impact on the late-time asymptotics of the decay
rate~\cite{Kehr-review,Hugues,Donsker-Varadhan}. Our analytical method
is based on a scheme that we name the `decoupled-ring'
approximation. Similar approaches have been used in the literature
\cite{WeiHav,Dombetal}.  We demonstrate that the late-time resummation
of our analytical result reproduces the known asymptotic, very slow
stretched-exponential
decay~\cite{Balagurov-Vaks,Donsker-Varadhan,Nieuwenhuizen2}. Section~\ref{1D-MC-sims}
investigates, by means of Monte Carlo simulations, the range of
validity for this analytic result when the trapping probability
departs from one, the value for which the decoupled-ring approximation
becomes exact. In Sec.~\ref{3D-MC-sims}, we examine, by way of
additional simulations, how the results presented resist to a
relaxation of the one-dimensional constraint, by allowing particles to
slightly diffuse also in the two transverse directions. Finally, in
Sec.~\ref{Experim} we discuss the possibility of applying such a model
to account for experimental measurements of exciton decay in organic
semiconductors, and comment on the crucial role of the spatial trap
distribution for the long-time kinetics.


\section{The model}
\label{Model}

We consider a one-dimensional, regular, infinite lattice ${\mathcal L}$
with linear lattice spacing $a$. A fraction $c$ of the lattice sites are
tagged, thereby denoting that they possess a special property. These
particular sites will hereafter be referred to as `traps', indicating
their specific role in the dynamics we are about to introduce.

In general, the spatial trap distribution can be arbitrary.
We shall single out two generic, contrasting situations:
\begin{enumerate}
    \item The trap distribution is {\em random}, i.e., the distance
    between two consecutive traps along the chain is represented by a
    stochastic variable with poissonian distribution.
\item The location of the traps is {\em spatially correlated}. This case
    typically covers the situation of $n$-modal intertrap spacings, or,
    more generically, situations where some kind of regularity
    ({\em order}) can be identified in the way traps are distributed
    over the system.
\end{enumerate}
Both situations are encountered in experimental samples:
If properly synthesized, semiconducting polymer chains in solution may
be considered to be essentially trap-free, except for their extremities
perhaps, thus falling into the second category. Photo-oxidation of the
same material leads to the creation of defects at random locations along
the chain, and the first of the two above cases then applies.

We now introduce entities, representing (pseudo- or quasi-)particles,
initially randomly distributed over the lattice sites, with density
$0<\rho_0\leq 1$. A discrete dynamics is implemented, by allowing
particles to perform symmetric random walks: A particle located at
site $k$ at time $t$ can subsequently be found either at $(k-1,t+\tau)$
with probability $1/2$, or at $(k+1,t+\tau)$ with the same probability,
where $\tau$ denotes the physical timestep associated to the iteration
of the dynamics.

At each timestep, every diffusing particle may annihilate through two
different decay channels:
\begin{itemize}
\item Spontaneous {\em decay}, occuring with fixed probability
      $0\leq p<1$. For excitons in organic semiconductors, this channel
      represents {\em radiative} recombination.
\item Stochastic {\em capture} by a trap: If the diffusing particle
      enters a tagged site, it annihilates with a given probability
      $0<q\leq 1$. For our later application to exciton kinetics, we
      specify this trap-mediated decay process to be {\em non-radiative},
      i.e., not emitting light at visible wavelengths.
\end{itemize}
In both cases, the particle is simply removed from the system. The
situation with $q=1$ will be referred to as {\em perfect traps}.

As both decay process act concurrently without directly influencing
each other, their combined effect appears as the product of two
{\em independent} factors. The spontaneous decay being trivially
described by a simple exponential relaxation, the real difficulty here
rests in the contribution to the kinetics related to the traps, where
the dynamical build-up of spatio-temporal correlations renders any
mean-field type of approach invalid in (sufficiently) low dimensions.
As we shall see, the spatial arrangement of the traps may crucially
determine the long-time asymptotics of the particle decay in this
situation.


\section{Asymptotic kinetics in the decoupled-ring approximation}
\label{RingDecSol}

\subsection{Motivations}

\begin{figure}[htb]
\centerline{\epsfig{file=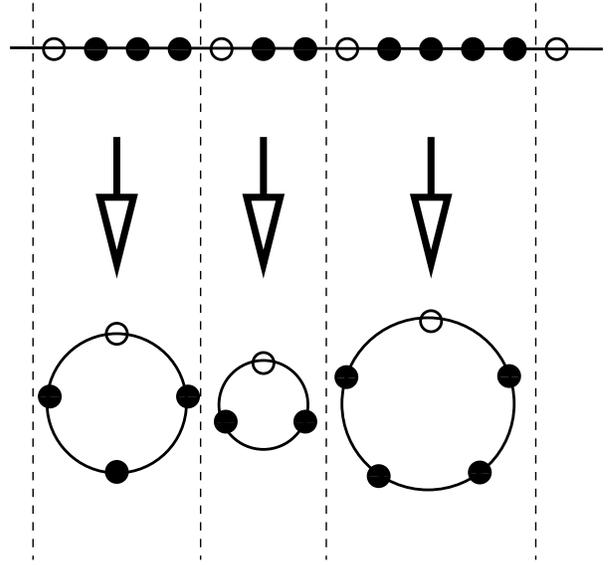,width=8cm}}
\caption{Schematic illustration of the {\em ring-decoupling procedure}.
Black dots represent regular lattice sites, white ones those with a
trap. The infinite, one-dimensional chain is split into segments
(delimited by the location of the traps), each of which is then closed
onto itself to form a ring. The entire chain is thereby mapped onto an
infinite set of rings of varying lengths. This approximation becomes
exact in the limit of perfect traps.}
\label{LineToRings}
\end{figure}

The model of perfect, uncorrelated traps in one dimension has been
solved by Anlauf \cite{Anlauf} by means of the span distribution
function for one-dimensional random walks.  The model of imperfect
trapping has been treated in the literature by various means.  Weiss
and Havlin \cite {WeiHav} studied the system in one dimension by
introducing a modified model where particles are either destroyed or
\textit{reflected} at a trap, thus decoupling the line segments
separated by a trap. This calculation was redone and some errors were
corrected in \cite{Nieuwenhuizen2}. Additionally, the coupled system
was mapped to a harmonic chain with random masses and the authors were
able to obtain the long-time behaviour of the particle decay in one
dimension. We chose a similar approach to \cite{WeiHav} by decoupling
line segments by cutting the infinite line at each trap and bending
each segment onto itself to form a ring (see
Fig.~\ref{LineToRings}). The advantage of this method over the
approach of mapping to a harmonic system with random masses is that it
is easy to obtain results for correlated traps: since each decoupled
ring can be treated seperately, the final result is simply obtained by
averaging over the ring lengths with the appropriate probability
distribution. The disadvantage is that it contains uncontrolled
approximations since particles can never leave their ring and wander
off into other regions as they could in the original model. However,
computer simulations show that these approximations are surprisingly
good, in particular for not too low trapping probabilities.  For
perfect traps, the decoupled model even becomes exact since the traps
prevent passage between segments even in the original model.

\subsection{Single-ring solution}

Here, we briefly sketch the solution of the model on a single ring of length
$n$.  The discrete time dynamics takes the form of a master equation for the
particle density distribution vector ${\boldsymbol{\rho}}(t)\equiv
(\rho_1(t),\dots, \rho_n(t))$, with the trap located at site 1 without loss
of generality,
\begin{equation}
\label{mastereq}
    \left(
  \begin{array}{c}
    \rho_1 \\
    \rho_2 \\
    \vdots \\
    \rho_{n-1} \\
    \rho_n
  \end{array}
  \right)\!(t+\tau)
  =
  \left(
  \begin{array}{cccccc}
    0 & \frac{1}{2} & 0 & \cdots & 0 & \frac{1}{2} \\
    \frac{1-q}{2} & 0 & \frac 12 & & & 0 \\
    0 & \frac{1}{2} &  &  & & \vdots \\
    \vdots & &  & \ddots &  & 0 \\
    0 & & &  &  & \frac{1}{2} \\
    \frac{1-q}{2} & 0 & \cdots & 0 & \frac{1}{2} & 0
  \end{array}
  \right) \!
  \left(
  \begin{array}{c}
    \rho_1 \\
    \rho_2 \\
    \vdots \\
    \rho_{n-1} \\
    \rho_n
  \end{array}
  \right)\!(t) \, .
\end{equation}

For late times, the overall decay rate of the concentration is `slaved' to the
largest eigenvalue $\alpha_n$ of the above matrix, and its corresponding
eigenvector $\boldsymbol{a}$. The calculation is straightforward and is
shown in App.~\ref{eigenvecval}; the result is
\begin{align}
\alpha_n &= \cos\phi_n \text{, where}\\
\phi_n &=  {\pi \over n}-{2\pi(1-q) \over q}{1 \over n^2}+
  {4\pi(1-q)^2 \over q^2}{1 \over n^3}+{\mathcal O}(n^{-4})
\end{align}
for the eigenvalue and
\begin{align}
a_1 &= 1, \\
a_k &= \cos(k-1)\phi_n + q \frac{\sin(k-2)\phi_n}{\sin\phi_n}
\end{align}
for the eigenvector.

Starting from an initial distribution $\rho_k(t=0)$, the
long-time solution is
\begin{align}
\boldsymbol{\rho}(t) &\stackrel{t,n \to \infty}{\sim}
  \frac{(\boldsymbol{a}, \boldsymbol{\rho}(0))}
  {(\boldsymbol{a}, \boldsymbol{a})}\,
  \boldsymbol{a}\,e^{t \ln \alpha_n},
\end{align}
where $(\cdot,\cdot)$ denotes the scalar product. The scalar products can be
worked out for a homogeneous initial distribution, $\rho_k(0)=\rho_0$, to
leading order in $1/n$, and the result for the mean density on the ring
$\overline{\rho}$ is
\begin{align}
\overline{\rho}(t) \equiv \frac 1n \sum_{k=1}^{n}\rho_k(t) &\stackrel{t,n \to
\infty}{\sim} \frac {8}{\pi^2}\rho_0\, e^{t\ln\alpha_n}.
\label{nresult}
\end{align}
Note that this result is better than might be expected at first sight because
the overlap of the eigenvector $\boldsymbol{a}$ with the initial homogeneous
configuration is very large (see App.~\ref{eigenvecval}),
\begin{align}
\frac{(\boldsymbol{a},\boldsymbol{\rho}(0))}
{|\boldsymbol{a}||\boldsymbol{\rho}(0)|}
&= \frac{2\sqrt{2}}{\pi} + \mathcal{O}(1/n) \approx 0.9\dots,
\end{align}
which guarantees that this result is good even at intermediate times.

\subsection{Full-chain solution}

Given an {\em arbitrary} distribution of traps, one can now write down
the solution for the full chain (in the decoupled-ring approximation)
by simply superposing the single-ring results, properly weighted with
the relative abundance for the occurrence of rings of length $n$ in
the chain. For example, if there are well-defined trap spacings such
that only a {\em finite} number of certain values of $n_i, i=1,\ldots,k$
is possible, with relative abundance $P(n_i)$, then the final result is
simply a sum of $k$ terms of the form (\ref{nresult}), each multiplied
with $P(n_i)$. As is evident from Eq.~(\ref{nresult}), the terms for
large $n$ will quickly dominate, and the maximum value $n_{\rm max}$
will govern the long-time limit.

For a random (poissonian) initial distribution of the traps with concentration
$0<c<1$, all values of $n$ are possible, the probability of finding a ring of
size $n$ being given by
\begin{equation}
  P(n) = n c^2 (1-c)^{n-1} \ .
\end{equation}
This allows us to finally write
\begin{eqnarray}
  &&\overline{\rho}(t) \stackrel{t \to \infty}{\sim}
  \frac{8}{\pi^2} \, \rho_0 \, c^2 \sum_{n=1}^{\infty} n(1-c)^{n-1}
  \times \label{result} \\
  &&\times \exp \left[ t \ln\cos \left(
  {\pi \over n}-{2\pi(1-q) \over q}{1 \over n^2}+
  {4\pi(1-q)^2 \over q^2}{1 \over n^3} \right) \right] . \nonumber
\end{eqnarray}

For $t \to \infty$, the late-time asymptotics of the {\em infinite} sum
(\ref{result}) can be extracted (see the appendix). The result reads:
\begin{eqnarray}
    \overline{\rho}(t) &\stackrel{t \to \infty}{\sim}& \rho_0 \,
    \frac{8}{-\ln(1-c)} \, \frac{c^2}{1-c} \, \sqrt{\frac{2}{3\pi}} \,
    t^{1/2} \times \nonumber \\
    &&\times \exp \left[-\frac{2(1-q)}{q}\ln(1-c)\right] \times \nonumber \\
    &&\times \exp \left[ -\frac 32 \left[-\pi\ln(1-c)\right]^{2/3}t^{1/3}
    \right] \ .
  \label{stretchedexp-Timo}
\end{eqnarray}

We recover the known {\em stretched exponential} long-time
decay~\cite{Donsker-Varadhan,Nieuwenhuizen2,Anlauf}
$\overline{\rho}(t)\sim\exp (-\text{const.}\,t^{1/3})$
with the correct values for both the exponent and its prefactor, and including
a $q$-dependent enhancement factor in agreement with
\cite{Nieuwenhuizen2} (higher order correction terms have been omitted here
for brevity). With a {\em random} (Poissonian) distribution of traps,
one therefore expects a much {\em slower} asymptotic time decay than for a
typical correlated distribution. Certainly, for any finite number of allowed
values of $n$, the asymptotic temporal decay will be a simple exponential. As
another example, a Gaussian distribution of the lengths of trap-free regions
(around some mean value) leads to a stretched exponential with exponent $1/2$
(by a calculation analogous to the one above). Thus, in addition to the trap
concentration and the induced decay probability $q$, the spatial arrangement
of the traps is of crucial importance.

Instead of using the asymptotic result Eq.~(\ref{stretchedexp-Timo}) we prefer
to use Eq.~(\ref{result}) (suitably truncating the sum) for our comparisons
with simulations and experiments below since it applies not only to the long
times but also to intermediate times. This is essential as it is known that
the crossover to the long-time behaviour Eq.~(\ref{stretchedexp-Timo}) may set
in so late to be unreachable in practice (see e.g.\ \cite{Barkema} and
references therein, \cite{Nieuwenhuizen1,Gallos}).


\section{Monte Carlo simulations}

\subsection{Strictly one-dimensional system}
\label{1D-MC-sims}

We have investigated the range of validity of the result (\ref{result})
emerging from our decoupled-ring approximation by means of Monte Carlo (MC)
simulations. The parameter of interest is of course $q$, the trapping
probability. For $q$ equal (or very close) to $1$, Eq.~(\ref{result}) should
perform very well. This agreement may be expected to degrade when $q$ is
lowered significantly below unity, since this implies an increasing coupling
between adjacent rings that has been completely neglected in our treatment of
the problem.

\begin{figure}[htb]
\vspace{0.5cm}
\centerline{\epsfig{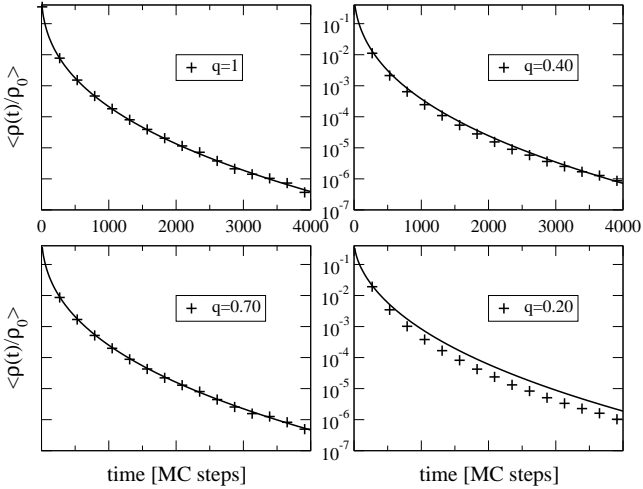}}
\caption{Monte Carlo simulations for the one-dimensional imperfect
trapping process. The dimensionless quantity on the vertical axis is
the amount of random walkers present in the system, normalized by the
initial total population. It can also be interpreted in a statistical
sense, as the survival probability of a {\em single} walker. Crosses
represent the numerical result, averaged over $\sim 5000$ independent
runs. The solid lines represent the analytical result of the
ring-decoupling scheme, as given by Eq.~(\ref{result}). The following
choice of parameters was made: system size=$2^{14}=16384$ sites; $p$
(spontaneous decay probability) $=0$; $c$ (trap concentration) $=0.2$;
diffusion coefficient $=0.5\;[a^2/\tau]$; $\rho_0$ (initial particle
density) $=0.6$.}
\label{1D-MC-sims-fig1}
\end{figure}
Figure~\ref{1D-MC-sims-fig1} shows a set of plots comparing results of
MC simulations to the approximate analytic formula (\ref{result}), for
different values of the trapping probability $q$. The spatial trap
distribution is random (poissonian). For clarity, we have set the
spontaneous decay probability $p$ to zero here. No crossover may anyway
be expected between the exponential and stretched exponential components
of the dynamics: Due to the structure of Eq.~(\ref{stretchedexp-Timo}),
the sub-exponential behavior is screened out by the spontaneous
exponential decay for all real and positive values of the time $t$.

The first (upper left) graph for $q=1$ serves as a reference for a
visual evaluation of the uncertainty in our numerical data. Looking at
the other graphs in the figure, one observes that for the chosen trap
concentration ($c=0.2$), the analytical result performs remarkably well
down to $q \cong 0.4$ (!). Note that the finiteness of the systems
simulated implies a temporal horizon: Beyond this limit, both strong
depletion in particles and inaccurate statistics for large, trap-free
regions (which play an increasingly important role for late times)
conspire to render the simulation meaningless. Our approximation being
an asymptotic theory, there is however no doubt that the matching
between Eq.~(\ref{result}) and the numerical data is at least as good
for the unexplored, late time domain as it appears in the `early-time'
regime depicted in Fig.~\ref{1D-MC-sims-fig1}.

\begin{figure}[htb]
\vspace{0.5cm}
\centerline{\epsfig{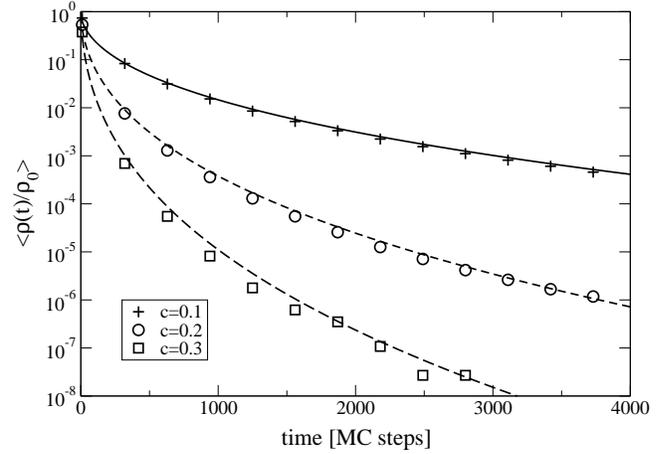}}
\caption{Monte Carlo simulations for the one-dimensional imperfect
trapping process with fixed trapping probability and varying traps
concentration.
The quantities on the axis and the meaning of the symbols chosen
are the same as for figure~\ref{1D-MC-sims-fig1}. Results have
been averaged over $\sim 5000$ independent runs. The following
choice of parameters was made: system size $=2^{14}=16384$ sites;
$p$ (spontaneous decay probability) $=0$; $q$ (trapping
probability) $=0.4$; diffusion coefficient $=0.5\;[a^2/\tau]$;
$\rho_0$ (initial particle density) $=0.6$.}
\label{1D-MC-sims-fig2}
\end{figure}

Figure~\ref{1D-MC-sims-fig2} illustrates how the decoupled-ring
approximation performs upon varying the trap concentration at
fixed trapping probability.

\subsection{Anisotropic three-dimensional system}
\label{3D-MC-sims}

Semiconducting polymer chains can be found in various configurations:
When in solution, they can be considered as totally isolated from
each other, hence representing a set of one-dimensional segments on
which excitons propagate and interact. They can, however, also be
organized in a more cohesive manner, namely in the form of a network
characterized by a certain degree of coupling between the polymer
chains, or even in highly ordered structures such as polycrystalline
films. This interaction between chains is evident both in the
chain-to-chain transfer of charges when these networks are used in
light emitting diodes, as well as in the chain-to-chain transfer of
excitons~\cite{Halls}. In these cases, the one-dimensional character
of the exciton dynamics is likely to be broken, the `particles' being
able to `cross-diffuse' from one chain to a neighboring one.

In this context, an interesting question that naturally arises is the
following: To which extent does the picture related to the ideally
one-dimensional case break down, once a small degree of cross-diffusion is
allowed in a set of one-dimensional chains forming a crystal-like structure?
For we know that in an isotropic three-dimensional system with uncorrelated
traps, the long-time behaviour of the system is given by a stretched
exponential of the form $\rho(t)\sim \exp(-\text{const.}\,t^{3/5})$
\cite{Balagurov-Vaks,Donsker-Varadhan}.

Our one-dimensional model is extended to three dimensions by allowing
particles to hop in all three spatial directions, but with different
rates. In particular, we single out one direction as the direction of
the polymer chains and allow hopping along this direction as in the
one-dimensional model with a diffusion constant $D_l$. Transverse
hopping is suppressed by choosing a transverse diffusion constant
$D_t<D_l$. Thus, a particle hops along the polymer direction with
probability $P_l=(2D_t/D_l+1)^{-1}$ and perpendicular to it with
probability $1-P_l$ at each time step.

\begin{figure}[htb]
\vspace{0.5cm}
\centerline{
        \epsfxsize=8.5cm
        \epsfbox{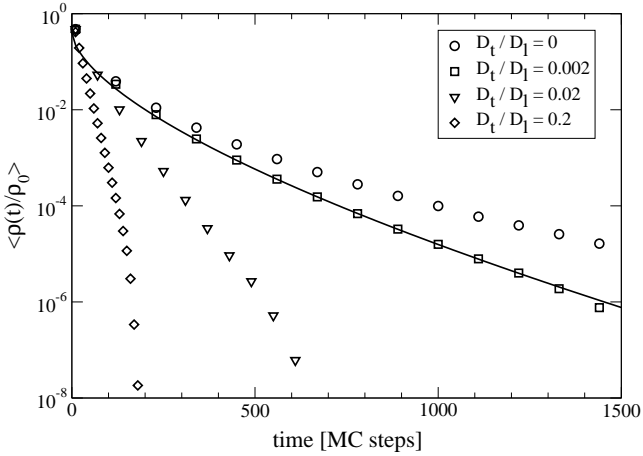}
        }
\caption{Monte Carlo simulations of a three-dimensional imperfect
trapping process competing with spontaneous decay (see the caption of
Fig.~\ref{1D-MC-sims-fig1} for comments on the axis and units).
Symbols represent the numerical results for different values of the
transverse diffusion coefficient $D_t$, as indicated in the
legend. They have been averaged over ca.\ 100 runs. The solid line
shows a fit with a stretched exponential $\exp(-a t^{3/5})$ as
appropriate for a truly three-dimensional system, showing that at
$D_t/D_l=0.002$ and $c=0.2$ the system already behaves like a
three-dimensional one.  The following choice of parameters was made:
system size $=2^{12}\times 2^{5}\times 2^{5}$ sites; $p=0.001$;
$c=0.2$; $q=0.5$; longitudinal diffusion coefficient
$D_l=0.5\;[a^2/\tau]$; $\rho_0=0.6$.}
\label{3D-MC-sims-fig1}
\end{figure}

Figure~\ref{3D-MC-sims-fig1} displays the result of MC simulations of
our model in three dimensions, where the ratio of the transverse and
longitudinal diffusion rates $D_t/D_l$ has been chosen non-zero, but
kept small. The virtue of this figure is to undoubtedly show the extent
to which the topology affects the dynamics in this process: Even a very
small relaxation of the one-dimensional constraint results in a drastic
effect. This is easily understood if one takes into consideration the
two following important aspects of the dynamics:
\begin{itemize}
\item The increasing role played by trap-free regions of larger and
larger size as time increases.
\item The fact that a particle jumping transversely onto a neighboring
chain has, statistically, large chances to penetrate a small inter-trap
segment, and therefore decay rapidly. This is of course true only under
the hypothesis that the trap distributions for all chains are totally
uncorrelated with respect to each other.
\end{itemize}
On the other hand, we see that the decoupled-ring approximation, which
correctly captures only the one-dimen\-sional topology, numerically
remains a fairly adequate description for small anisotropy ratios.

\begin{figure}[htb]
\vspace{0.5cm}
\centerline{
        \epsfxsize=8.5cm
        \epsfbox{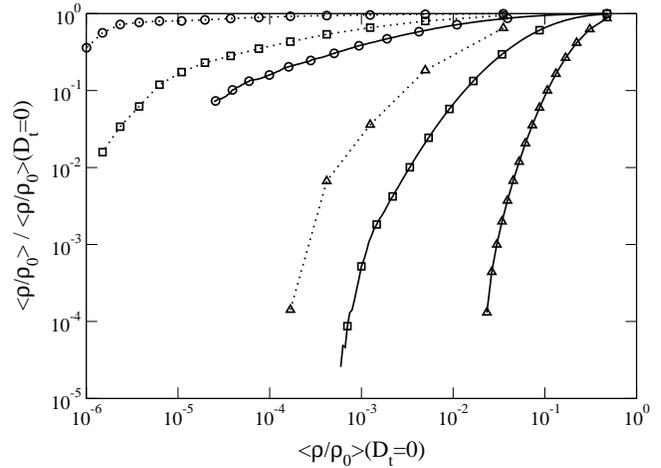}
           }
\caption{Effect of the trap concentration on the relaxation of
the one-dimensional constraint. The same simulations as in
Fig.~\ref{3D-MC-sims-fig1} were performed for a higher trap
concentration $c=0.7$ (averaged over $\sim$ 100 runs here, too), and
both sets are now displayed together. Solid lines correspond to
$c=0.2$, and dotted ones to $c=0.7$. For each of the three cases with
$D_t \neq 0$, the plot shows the {\em relative} deviation from the
$D_t=0$ case. Circles, squares, and triangles indicate respectively
the simulations with values $D_t/D_l=0.002,0.02$, and $0.2$.}
\label{3D-MC-sims-fig2}
\end{figure}

In Fig.~\ref{3D-MC-sims-fig2}, we examine how the trap concentration
affects the amplitude of the effect of the relaxation of the
one-dimensional constraint on the population decay. As shown, a higher
trap concentration helps to screen out the emergence of the transverse
channel for diffusion. This is an important observation when one tries
to use our model to interpret experimental data, as we will see in the
forthcoming section.


\section{Applicability to the decay of excitons in quasi-one-dimensional
     organic semiconductors}
\label{Experim}

As previously mentioned, our model may be considered a potential candidate
for a `minimal' description of the decay dynamics of excitons in organic
semiconductors. Confrontation with real data coming from experimental
measurements of such a decay is therefore instructive. The figures below
display our attempts to fit such experimental data with our analytical
result~(\ref{result}), supplemented with a spontaneous decay factor
$e^{-pt}$.

The data presented here, orginally reported in Ref.~\cite{Wolle},
measure (in frequency space) the response of ladder-type
Poly(\textit{Para}-Phenylene) (LPPP) to excitation by a modulated
light beam. The signal represents the Fourier transform of the
spectrally integrated photoluminescent emission of the material. This
is given by the {\em activity}, i.e., the number of {\em radiative}
recombinations of excited electronic states (excitons) per time
interval. Thus, while both spontaneous and trap-induced decay channels
reduce the exciton population, only the activity due to the
spontaneous radiative recombination is monitored by the data. Three
different types of samples, all made from the same material, were
analyzed: The polymer in solution, a pristine polymer film, and the
same film after it had been subjected to photooxidization. In the
following, we report our comparison of theoretical and experimental
data for each case.

\begin{figure}[htb]
\centerline{
        \epsfxsize=8.5cm
        \epsfbox{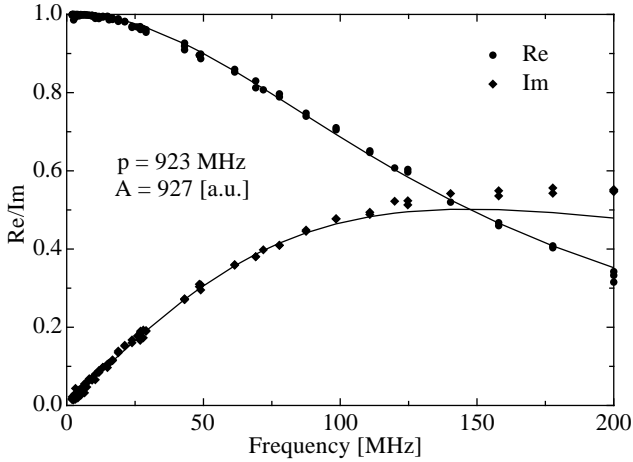}
           }
\caption{Fourier transform of the response function as measured (dots)
and a fit (solid lines) with the Fourier transform of the activity for
an exponential time decay for the polymer in solution. Fit parameters
are the spontaneous decay rate $p$ and an (irrelevant) overall amplitude
$A$, given in arbitrary units.}
\label{Solution-fig}
\end{figure}

\begin{figure}[htb]
\centerline{
        \epsfxsize=8.5cm
        \epsfbox{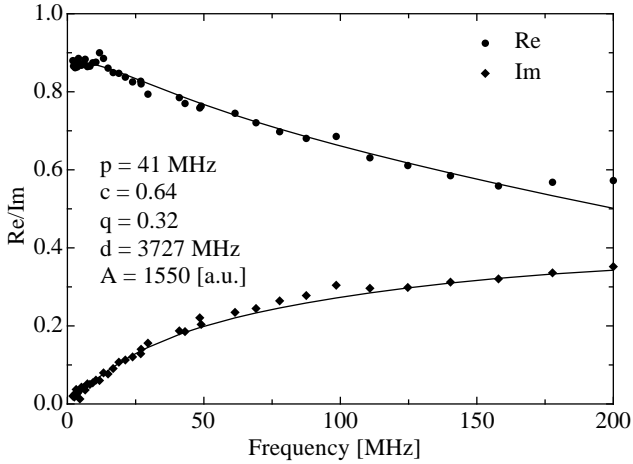}
           }
\caption{Fourier transform of the response function as measured (dots)
and a fit (solid lines) with the Fourier transform of the activity
corresponding to the time decay according to Eq.~(\ref{result}) for the
photo-oxidized film. Fit parameters are the spontaneous decay rate $p$,
the trap concentration $c$, the trapping probability $q$, the exciton
hopping rate $d$, and an (irrelevant) overall amplitude $A$. The
agreement is quite satisfactory.}
\label{Filmox-fig}
\end{figure}

\begin{figure}[htb]
\centerline{
        \epsfxsize=8.5cm
        \epsfbox{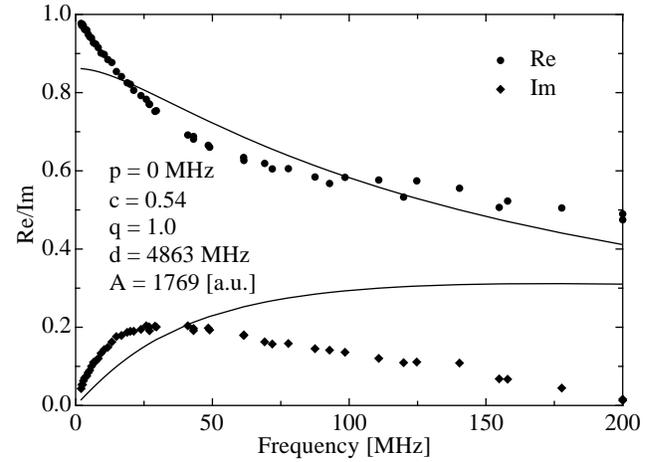}
           }
\caption{Fourier transform of the response function as measured (dots)
and a fit (solid lines) with the Fourier transform of the activity
corresponding to the time decay according to Eq.~(\ref{result}) for the
non-oxidized film. Fit parameters are as in Fig.~\ref{Filmox-fig}. The
agreement is very poor, see text.}
\label{Film-fig}
\end{figure}

Figures~\ref{Solution-fig}--\ref{Film-fig} show the experimental data
as real and imaginary parts. The solid lines are simultaneous fits
(performed with a Levenberg-Marquardt algorithm) to both real and
imaginary parts with a Fourier transform of the activity corresponding
to a single exponential in the case of the solution,
Fig.~\ref{Solution-fig}, and to a decay following Eq.~(\ref{result})
in the case of the photo-oxidized and non-odixized films,
respectively, Figs.~\ref{Filmox-fig} and \ref{Film-fig}. 

For the solution, Fig.~\ref{Solution-fig}, the experimental data can
very well be described by a single exponential (Lorentzian in Fourier
space). The polymer chains have very few defects, apart perhaps from
their end-points which may act as traps, and have all the same lengths
--- this results in an almost purely exponential decay of the exciton
population since the lengths of the trap-free regions are monodisperse
(in the sense of the decoupled-ring approximation).

For the oxidized film, Fig.~\ref{Filmox-fig}, the situation changes
entirely. First, when going from solution to film we introduce a
coupling between the chains which is {\em not} present in the
solution. Second, due to oxidization, many defects acting as traps
have been introduced on the film, see Ref.~\cite{Wolle}(b).
Accordingly, Eq.~(\ref{result}) describes the data well, even though
it has been devised for one-dimensional systems rather than films.
This may be explained by the large number of traps (around 65\% of the
{\em effective} sites according to the fit), which prevent the
excitons from `feeling' the three-dimensional nature of the film, as
Fig.~\ref{3D-MC-sims-fig2} clearly illustrated in the preceding
section. Further support for this effective confinement of excitons to
a single chain and the high number of traps is given in
Ref.~\cite{Wolle}(b), based on an analysis of the emission and
absorption photoluminescence spectra. 

There are, of course, errors in the values of the fit parameters. The
greatest uncertainty comes from the spontaneous decay rate $p$, which
induces very large errorbars on all other parameters as well (not
shown in the figures). If, however, $p$ is kept fixed at the value
indicated in Fig.~\ref{Filmox-fig} and only the remaining parameters
are used for fitting, the error in e.g. the trap concentration is
found to be about 15\%. Nevertheless it should be kept in mind that
our most important result here is the functional form of our
theoretical curve which is able to reproduce the data rather than
precise values of the parameters.

The observed photoluminescence decay-time has been greatly {\em
increased} compared to the solution, in spite of the presence of
traps. At the same time we know that the major decay path for this
sample is non-radiative because the photoluminescence quantum yield is
only a few percent. Hence, the long-living photoluminescence must be
the signature of a `stabilized species' which is {\em not}
photogenerated with a high yield. This may be due to low-energy sites
on the polymers which capture the excitons and prevent them from
decaying spontaneously, or delayed photoluminescence due to
triplet-triplet annihilation. Evidence for the latter process in this
particular material can be found in Ref.~\cite{rom}. Neither of these
processes have been included in our simplified model.

Turning to the last of the three sets of data, the non-oxidized film
shown in Fig.~\ref{Film-fig}, we observe that it can neither be
described by a single exponential nor with Eq.~(\ref{result}).
Fig.~\ref{Film-fig} shows the result of an unsuccessful fitting
attempt with Eq.~(\ref{result}). Apparently, the influence of the
three-dimen\-sional nature of the film made from the {\em same}
molecules as were measured in solution first, cannot be described
simply by the introduction of traps. Accordingly, since our model is 
not appropriate for the non-oxidized film, the actual values of the fit 
parameters as shown in Fig.~\ref{Film-fig} are entirely meaningless.

This has an important implication. The non-oxidized film has a
photoluminescence quantum yield of 30\% vs. nearly 100\% for the
solution of the same molecules. One way to account for this could be
the formation of traps during the film forming process, e.g. by
conformational stress on the molecules~\cite{TQ}. If this was the case,
our model would capture this effect and describe it properly. Hence, the
reduction of the quantum yield and the difference in photoluminescence
dynamics between film and solution have to originate from a
{\em different} solid state effect.


\section{Summary and conclusion}

We have considered a model of random walkers undergoing decay through
both capture by imperfect traps {\em and} spontaneous decay. In one
dimension, we have derived an explicit analytical form for the time
dependence of the survival probability, based on the ring-decoupling
approximation scheme, supplemented with an asymptotic expansion. We have
demonstrated that this result, although exact only in the limit of
perfect traps, performs remarkably well upon lowering of the trapping
probability when confronted to numerical simulations. Extension of our
simulations to three dimensions, albeit with anisotropic diffusion,
have illustrated the high sensitivity of the decay rate to the
low-dimensionality constraint.

An application of our findings to the understanding of the decay dynamics
of excitons in semiconducting polymers has been attempted. We conclude
that within our model traps alone cannot account for the observed
difference between exciton dynamics in a pristine solution and film made
from the same conjugated polymer, LPPP. Photo-oxidizing LPPP leads to
isolated segments which probably correspond to the one-dimensional case
in our model. The large concentration of effective trap sites obviously
dominates over all other physically relevant mechanisms.

In this context, it is important to emphasize again the crucial role
played by the {\em spatial} trap distribution. For a more or less regular
spacing of defects, the long-time kinetics would be governed by a simple
exponential, with a `renormalized' decay rate. For a random trap
distribution, on the other hand, a much slower stretched exponential
decay ensues asymptotically. At least in the long-time limit, one might
therefore quite drastically control the exciton population through
appropriate `engineering' of the spatial arrangement of the trapping
defects.

At any rate, our model system demonstrates again the importance of
including statistical fluctuations effects effectively low-dimensional
samples. This is of course a well-established fact in statistical
mechanics model systems. Yet to date there have been few examples where
clear evidence of correlations not captured by mean-field theory has
been found in real experiments on non-equilibrium systems. Moreover, we
believe that our study underscores the relevance of simplified kinetic
models, at least in the long-time limit, where detailed microscopic
mechanism become less prominent. We hope that a better understanding
of, e.g., the important organic semiconducting materials will eventually
be achieved through combining experimental data with both
quantum-mechanical computations and more macroscopic, statistical
approaches.


\begin{acknowledgement}
J.M.\ is grateful to Bastien Chopard, head of the Scientific Parallel
Computing Group at the Department of Computer Science of the
University of Geneva, for his kind permission to use the Connection
Machine CM2, on which the simulations appearing in this paper have
been performed.  He also acknowledges support from the Swiss National
Science Foundation, under fellowship nr. 81GE-59927.  This work has
also been supported through grants from the National Science
Foundation (Grant no. DMR-0075725) and the Jeffress Memorial Trust
(Grant no. J-594). T.A.\ acknowledges support by the DFG under grant
Zi209/5-1 and by the German Academic Exchange Service (DAAD) under a
postdoc fellowship. We thank Beate Schmittmann and Massimiliano Di
Ventra for helpful discussions.
\end{acknowledgement}


\appendix

\section{Appendix: Decoupled-ring approximation}

\subsection{Computation of eigenvectors and eigenvalues}

\label{eigenvecval}

In this appendix, we show how to compute the {\em symmetric} eigenvectors
and eigenvalues of the matrix
\begin{equation}
  M = \left(
  \begin{array}{cccccc}
    0 & \frac{1}{2} & 0 & \cdots & 0 & \frac{1}{2} \\
    \frac{1-q}{2} & 0 & \frac 12 & & & 0 \\
    0 & \frac{1}{2} &  &  & & \vdots \\
    \vdots & &  & \ddots &  & 0 \\
    0 & & &  &  & \frac{1}{2} \\
    \frac{1-q}{2} & 0 & \cdots & 0 & \frac{1}{2} & 0
  \end{array}
  \right)
\end{equation}
appearing in Eq.~(\ref{mastereq}). The anti-symmetric eigenvectors,
also present in principle, are irrelevant in this context because
they have vanishing overlap with the initial homogeneous particle
distribution. We therefore try a symmetric ansatz of the form
\begin{equation}
{\bf a} \equiv
  \left(
  \begin{array}{c}
    1 \\
    a_2 \equiv \alpha \\
    a_3 \\
    a_4 \\
    \vdots \\
    a_{n-2} = a_4 \\
    a_{n-1} = a_3 \\
    a_n = a_2 = \alpha
  \end{array}
  \right) \ ,
\end{equation}
and note that this implies $(M\mathbf{a})_{1}=\alpha$ which identifies
$\alpha$ as the prospective eigenvalue corresponding to $\mathbf{a}$.
We can then read off that the second component of $M\mathbf{a}$ is
given by
\begin{equation}
    (M\mathbf{a})_{2} = \frac{1-q}{2}a_{1} + \frac 12 a_{3} \ ,
\end{equation}
which must be equal to $\alpha a_{2}$ since $\mathbf{a}$ is supposed
to be an eigenvector. This gives
\begin{equation}
    a_{3} = 2\alpha^2 - 1 + q \ .
\end{equation}
By a similar argument, the following recursion relation can be derived
for the other entries of the eigenvector:
\begin{equation}
    \label{recursion}
    a_{k} = 2\alpha a_{k-1} - a_{k-2}\quad (k\ge 4) \ .
\end{equation}
This kind of second-order recursion relation has two exponential
solutions $r_{\pm}^{k-2}$ (the $-2$ in the exponent is for later
convenience), and $a_{k}$ is given by a linear superposition of them,
$a_{k}=Ar_{+}^{k-2}+Br_{-}^{k-2}$. Plugging the ansatz $r^{k-2}$ into
the recursion relation Eq.~(\ref{recursion}) yields a quadratic
equation for $r$ with the two solutions $r_{\pm}=\alpha\pm
i\sqrt{1-\alpha^2}$. Noting that $0\le\alpha<1$ is implied by the
non-conserved number of particles owing to the trap and the fact that
the particle density cannot be negative, this can be written in terms
of an angle $\phi$ with $\alpha=\cos \phi$ as $r_{\pm}=e^{\pm i\phi}$.

The initial conditions $a_{2}=\alpha$ and $a_{3}=2\alpha^2 - 1 + q$ now
determine the constants $A$ and $B$; a straightforward calculation gives
\begin{equation}
    A=B^*=\frac 12 e^{i\phi} + \frac{q}{e^{i\phi}-e^{-i\phi}} \ ,
\end{equation}
resulting in
\begin{equation}
    \begin{split}
    a_{1}&=1 \ , \\
    \label{eigenvec}
    a_{k}&=\cos(k-1)\phi + q\frac{\sin(k-2)\phi}{\sin\phi}
\quad(2\le k\le n) \ .
    \end{split}
\end{equation}
It now only remains to be checked that $a_{n}=a_{2}$, as imposed from the
beginning. This condition determines $\phi$ and, after some algebra,
results in the following equation:
\begin{eqnarray}
    && [1-(1-2q)\cos 2\phi] \sin n\phi \nonumber\\
    &&\quad - [1-(1-2q)\cos n\phi ] \sin 2\phi = 0 \ .
    \label{eigenval}
\end{eqnarray}
The smallest non-zero solution to this equation leads to the sought-for
largest eigenvalue $\alpha=\cos \phi$ [even though $\phi=0$ is technically
a solution to Eq.~(\ref{eigenval}), it is not a valid one since the
constant $A$ does not exist for $\phi=0$].
Assuming that the smallest $\phi$ can be represented by a perturbative
expansion as $\phi=\frac{c_{1}}{n}+\frac{c_{2}}{n^2}+\cdots$,
Eq.~(\ref{eigenval}) can be expanded in powers of $1/n$ and the
coefficients $c_{i}$ are readily evaluated. The result is
\begin{equation}
    \phi = \frac{\pi}{n} - 2\pi\frac{1-q}{qn^2} +
    4\pi\frac{(1-q)^2}{q^2n^3} + \mathcal{O}(n^{-4}) \ .
\end{equation}

\subsection*{Overlap of the eigenvectors with the initial state}

As seen in Sec.~\ref{RingDecSol}, we also need to know the overlap of
the eigenvector found in the preceding subsection with the initial
homogeneous particle distribution, i.e., we need to calculate
\begin{equation}
    (\mathbf{a},\pmb{1}) = \sum_{k=1}^{n} a_{k} \ , \ \text{ and } \
    (\mathbf{a},\mathbf{a}) = \sum_{k=1}^{n} a_{k}^2 \ .
\end{equation}
The symbol $\pmb{1}$ here stands for the vector with all entries equal
to $1$.
Both sums can be evaluated tediously but straightforwardly by inserting
Eq.~(\ref{eigenvec}) and noting that both of them can be written in
terms of geometric series. The resulting expression may be evaluated
further for large $n$ by inserting $\phi=\pi/n+\mathcal{O}(n^{-2})$.
The results read
\begin{equation}
    (\mathbf{a},\pmb{1}) \stackrel{n \to \infty}{\sim}
    \frac{2qn^2}{\pi^2} \ , \  \text{ and } \quad
    (\mathbf{a},\mathbf{a}) = \stackrel{n \to \infty}{\sim}
    \frac{q^2n^3}{2\pi^2} \ .
\end{equation}
Altogether this gives
\begin{equation}
    \frac{(\mathbf{a},\pmb{1})^2}{(\mathbf{a},\mathbf{a})}
    \stackrel{n \to \infty}{\sim}
    \frac{8n}{\pi^2} \ .
\end{equation}

\subsection{Asymptotic evaluation of the long-time behavior}

Eq.~(\ref{result}), the main result of this calculation, can be
evaluated asymptotically for long times. Since
$\phi=\mathcal{O}(1/n)$, the long-time behavior is dominated by the
large-$n$ terms in the sum. Using
$\alpha=cos(\phi)=1-\pi^2/2n^2+2\pi^2(1-q)/qn^3+\mathcal{O}(n^{-4})$
and ignoring irrelevant prefactors for the moment, Eq.~(\ref{result})
can be written as
\begin{equation}
    \overline{\rho}(t) \propto \sum_{n=1}^{\infty} n \,
    \exp \left[ n\ln(1-c) - \frac{\pi^2}{2n^2} \, t 
    + \frac{2\pi^2(1-q)}{qn^3} \, t \right].
\end{equation}
By introducing $x\equiv t^{1/3}$, the sum on the r.h.s.\ can be recast
as a Riemannian sum approximating an integral in the variable $z=n/x$,
\begin{eqnarray}
    \overline{\rho}(t) & \propto &
    x^2\sum_{n=1}^{\infty} \frac 1x \frac nx
    e^{\frac{x^3}{n^3}\frac{2\pi^2(1-q)}{q}}
    e^{x[n\ln(1-c)/x - \frac{\pi^2x^2}{2n^2}]} \\
     & \stackrel{x \to \infty}{\sim} &
     x^2 \int_{0}^{\infty} dz\,z e^{2\pi^2(1-q)/z^3 q}
       e^{x[z\ln(1-c) - \pi^2/2z^2]} \ .
\end{eqnarray}
This integral may now be evaluated asymptotically by Laplace's method
 and finally yields, including all prefactors,
Eq.~(\ref{stretchedexp-Timo}).


\end{document}